\RequirePackage{fix-cm}
\documentclass{svjour3}                     
\smartqed  
\usepackage{url}
\usepackage{mathtools, cuted}
\usepackage{natbib}
\usepackage{hyperref}
\usepackage{graphicx}
\usepackage{amssymb}
\usepackage{amsmath}
\usepackage{balance}
\usepackage{mathtools}
\usepackage{lineno}

\usepackage[dvipsnames]{xcolor}

\usepackage{ragged2e}
\justifying
\balance

\begin{document}

\title{Prediction of Fluid Flow in Porous Media by Sparse Observations and Physics-Informed PointNet
}

\titlerunning{Prediction of Fluid Flow in Porous Media}        

\author{Ali Kashefi         \and
        Tapan Mukerji 
}


\institute{Ali Kashefi \at
              Department of Civil and Environmental Engineering, Stanford University, Stanford, CA 94305 \\
              \email{kashefi@stanford.edu}           
           \and
           Tapan Mukerji \at
              Department of Energy Science and Engineering, Stanford University, Stanford, CA 94305\\
              \email{mukerji@stanford.edu}
}

\date{Received: date / Accepted: date}


\maketitle

\section*{Highlights}
\begin{itemize}
\item[$\blacksquare$] For the first time, physics-informed PointNet is used for predicting fluid fields within porous media at pore scales by sparse observations.
\item[$\blacksquare$] The effects of spatial correlation lengths, noisy sensor data, pressure observations, and average spatial distance of sensors are investigated.
\item[$\blacksquare$] The robustness of the framework is shown through error analysis of predicted velocity and pressure fields, and computed permeability.
\end{itemize}

\begin{abstract}
We predict steady-state Stokes flow of fluids within porous media at pore scales using sparse point observations and a novel class of physics-informed neural networks, called “physics-informed PointNet” (PIPN). Taking the advantages of PIPN into account, three new features become available compared to physics-informed convolutional neural networks for porous medium applications. First, the input of PIPN is exclusively the pore spaces of porous media (rather than both the pore and grain spaces). This feature diminishes required computer memory. Second, PIPN represents the boundary of pore spaces smoothly and realistically (rather than pixel-wise representations). Third, spatial resolution can vary over the physical domain (rather than equally spaced resolutions). This feature enables users to reach an optimal resolution with a minimum computational cost. The performance of our framework is evaluated by the study of the influence of noisy sensor data, pressure observations, and spatial correlation length.


\end{abstract}


\section{Introduction and motivation}
\label{intro}
Since the late 2018s, deep learning schemes have become popular for the study of various aspects of porous media. Examples are rock image segmentation \citep{da2002physical,karimpouli2019segmentation,phan2021automatic,niu2020digital}; reconstruction and enhancement of rock image resolution \citep{da2019enhancing,liu2022multiscale,niu2021geometrical,wang2020boosting}; prediction of geometric characteristics such as porosity \citep{alqahtani2020machine,bordignon2019deep,graczyk2020predicting}; prediction of physical properties such as permeability \citep{graczyk2020predicting,hong2020rapid,kashefi2021porous,wu2018seeing,al2023effective}, effective diffusivity \citep{wu2018seeing}, wave propagation velocities \citep{karimpouli2019image}; and prediction of velocity and pressure fields of fluids within pore spaces \citep{alhubail2022extended,kamrava2021simulating,santos2020poreflow,lu2022comprehensive,tartakovsky2018learning,wang2021physics,wang2021ml}. Our focus in the current research is the last one - prediction of fluid velocities and pressure. According to the literature, three main approaches have been taken so far for predicting fluid flow fields in pore spaces of porous media using available deep learning algorithms.

The first approach falls in the category of supervised learning, where the deep learning frameworks used have no information concerning the physics describing fluid flow fields within porous media \citep{santos2020poreflow,wang2021ml}. In this approach, plentiful labeled data are generated using numerical solvers \textcolor{blue}{ (e.g., lattice Boltzmann, finite-element, finite-volume, and fast Fourier transform methods)} or collected from lab experiments for training deep learning frameworks to learn an end-to-end mapping from the geometry of a porous medium to its velocity or pressure fields. Convolutional neural networks (CNNs) are commonly used in this approach. The loss functions of these deep learning frameworks are designed based on the mismatch between values predicted by a neural network and training labeled data (i.e., ground truth), mathematically quantified in $L^2$ norm or other norms.

The second approach also falls in the category of supervised learning; however, the employed deep learning frameworks are ``guided'' by the problem physics \citep{kamrava2021simulating,wang2021physics}. More precisely, the loss function in this approach is similar to the one specified in the first approach with the difference that it is regularized by extra terms representing the $L^2$ norm (or other norms) of the residuals of partial differential equations (PDEs), governing the physics of fluid flow in porous media. Technically, the PDEs of interest are discretized by a finite difference scheme and then the corresponding stencils are enforced into non-trainable filters in the last layer of CNNs. Compared to the first approach, the second one is faster in convergence, more generalizable, and lastly requires ``relatively'' smaller labeled training data. One may refer to \citet{wang2021physics} and \citet{kamrava2021simulating} for a detailed discussion of this comparison.

Nevertheless with the success of these two approaches, they come with a few shortcomings. First, both require plentiful training data. This is while generating and collecting labeled data are computationally and experimentally expensive. Second, CNNs used in both approaches take the grain and pore spaces as the network inputs and outputs, whereas we exclusively seek for the network predictions only in the pore spaces of porous media where the fluid exists. The integration of the solid grain spaces into CNNs demands extra memory (i.e., RAM). Third, because these two approaches employ CNNs, labeled data used for training have to be in a Cartesian grid format with uniform spacing. This format might lead to unrealistic representation of the geometry of pore spaces, specifically near the pore-grain boundaries. Furthermore, if labeled data are generated using finite element/volume methods on unstructured grids, the generated data have to be interpolated on the uniform Cartesian grids. Such interpolations introduce errors to the training data, and consequently to the network prediction. Additionally, executing the interpolation requires extra effort even before training CNNs. Fourth, CNNs are technically designed for uniform resolutions. Thus, to increase the resolution of an area of interest in pore spaces, users of these two approaches have no choice except to increase the Cartesian grid resolutions everywhere in the domain. This inflexibility enforces high computational costs to the machine learning system of both approaches. Next, we discuss the third approach to resolving and improving the above mentioned limitations.

The third approach falls in the category of weakly supervised learning and is mainly constructed based on the idea of physics-informed neural networks (PINNs). PINNs were first proposed by \citet{raissi2019physics} for solving forward and inverse problems and its latter versions with various enhancements and extensions have been introduced such as fPINN \citep{pang2019fpinns}, nPINN \citep{pang2020npinns}, B-PINN \citep{yang2021b}, hp-VPINN \citep{kharazmi2021hp}, PPINN \citep{meng2020ppinn}, PIPN \citep{KASHEFI2022111510}, etc. In PINN-based methodology, the loss function is mainly the $L^2$ norm (or other norms) of the residuals of governing equations of fluid flow fields of porous media, regulated by the $L^2$ norm of mismatch between neural network predictions and sparse scattered labeled data. Through the lens of porous media, PINNs have been used but only for predicting Darcy's flow on field scale \citep{tartakovsky2018learning,alhubail2022extended}. In the current study, for the first time, we use the concept of physics-informed neural networks for the prediction of Stokes flow at pore scales in porous media. Specifically, we use PIPN \citep{KASHEFI2022111510} as an advanced version of PINNs. Remarkable advantages of PIPN have been addressed in detail by \citet{KASHEFI2022111510}. Using the PIPN technology, all the above mentioned shortcomings of the first two approaches are obviated. Based on the fundamental mathematics of PIPN, only sparse scattered labeled data are required. We exclusively discretize the pore space with a set of scattered point cloud, where the spatial density of point distribution can vary freely over the space. Such freedom allows users to represent the geometry of the pore space and its boundaries smoothly and realistically. Additionally, the PIPN \citep{KASHEFI2022111510} framework can conveniently be integrated with unstructured grids, and no data interpolation is needed. In the rest of this article, we illustrate these features practically. In addition, we explore some practical issues in applying PIPN such as i) the effect of the type of data available at the sparse sensors (with and without pressure data); ii) the effects of noisy sensor data; and iii) the impact of sparsity in the sensor locations.

\section{Physics-informed PointNet (PIPN) for Stokes flows in porous media}

\subsection{Governing equations and mathematical definition}

We first describe the partial differential equations governing the physics of fluid flow fields in porous media at pore scales. The conservation of mass and momentum of an incompressible steady creeping (Stokes) flow of a Newtonian fluid are respectively written as
\begin{eqnarray}
\label{Eq1}
\nabla \cdot \textbf{\textit{u}}=0 \textrm{ in } V,
\end{eqnarray}
\begin{eqnarray}
\label{Eq2}
\nabla p - \mu \Delta \textbf{\textit{u}} =\textbf{{0}} \textrm{ in } V,
\end{eqnarray}
where \textbf{\textit{u}} and $p$ indicate respectively the velocity vector and pressure of the fluid with the dynamic viscosity of $\mu$. The pore space of a porous medium is denoted by $V$. We further show the $x$ and $y$ components of the velocity vector by $u$ and $v$, respectively. \textcolor{blue}{Note that one may alternatively consider the full Navier-Stokes equations (see e.g., \citep{hassanizadeh1987high}).} The permeability ($\mathcal{K}$) of the porous media in the $x$ direction is computed as \citep{darcy1856fontaines,berg2014permeability,eshghinejadfard2016calculation}

\begin{equation}
\label{Eq3}
\mathcal{K} = -\frac{\mu \bar{U}}{\Delta p/L},
\end{equation}
where $\bar{U}$ is the average $x$ velocity over the entire space of porous media. The term of $\Delta p/L$ denotes the applied \textcolor{blue}{constant} pressure gradient over the length $L$ in the $x$ direction of the porous media.

Mathematically, our goal is to solve an inverse problem of the Stokes flow using PIPN. It can be described as follows: given no slip boundary condition on the wall boundaries and a set of sparse labeled data of the velocity and pressure fields at sensor locations, find the full velocity and pressure fields at inquiry points. Moreover, we compute the permeability of the porous media using the predicted velocity fields.

\subsection{Physics-informed PointNet (PIPN)}

\begin{figure*}
\label{FigS1}
\centering
\includegraphics[width=0.95\textwidth]{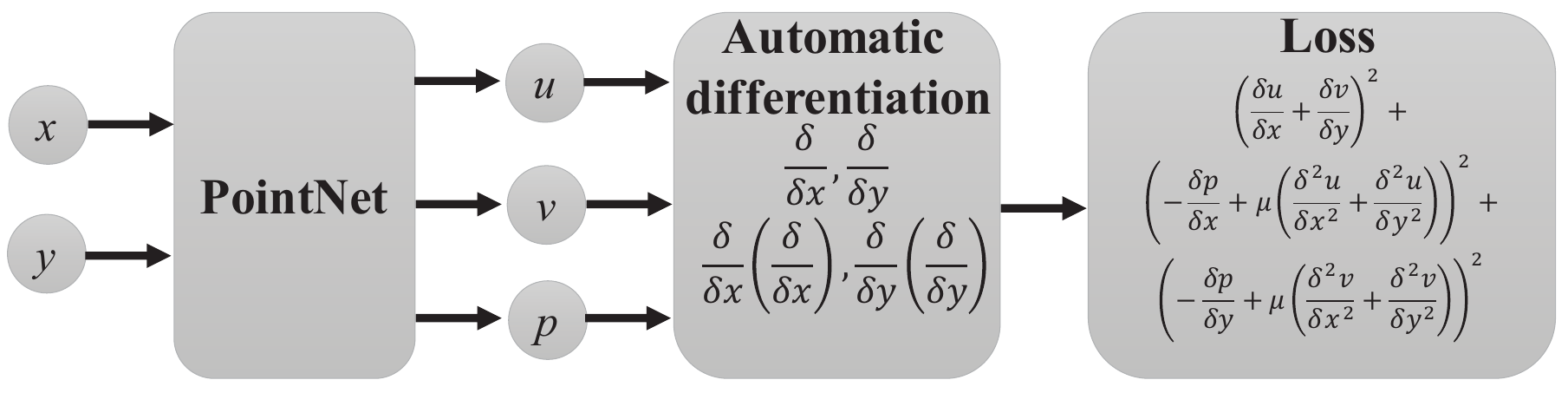}
\caption{Schematic flowchart of Physics-informed PointNet (PIPN) for Stokes flow in porous media. A full description of the loss function is presented in Eqs. (\ref{Eq4}--\ref{Eq10}).}
\end{figure*}

Historically, \citet{kashefi2021point} used PointNet \citep{qi2017pointnet} for the first time for supervised deep learning of incompressible flows on irregular geometries. The successes of applying PointNet \citep{qi2017pointnet} to the area of computational mechanics motivated \citet{KASHEFI2022111510} for proposing physics-informed PointNet (PIPN), which is a weakly supervised deep learning framework for incompressible flows. The PIPN methodology has been expressed in detail by \citet{KASHEFI2022111510}. Here we illustrate the PIPN framework from a general point of view and explain how to specialize it for the porous medium applications. Figure 1 depicts the general flowchart of PIPN. Accordingly, the space of a porous medium ($V$) is represented by $N$ points such that each point has the spatial $x$ and $y$ coordinates. In the next stage, we feed the constructed point clouds into PointNet \citep{qi2017pointnet}, while the outputs of PointNet \citep{qi2017pointnet} are the velocity ($u$,$v$) and pressure ($p$) values at the corresponding input points. Afterwards, we take spatial derivatives of outputs ($u$,$v$,$p$) with respect to the corresponding inputs ($x$,$y$) using the automatic differentiation technology of TensorFlow \citep{abadi2016tensorflow}. Finally, we build up the loss function of PIPN as the summation of residuals of governing equations (Eqs. \ref{Eq1}--\ref{Eq2}) as well as the mismatch between predicted outputs and sparse labeled data, all quantified in $L^2$ norm. Similar to the original version of PIPN proposed by \citet{KASHEFI2022111510}, we use the hyperbolic tangent activation function defined as

\begin{equation}
\label{S4}
\tanh(\lambda)=\frac{\exp(2\lambda)-1}{\exp(2\lambda)+1},
\end{equation}
in all the layers of PIPN. Note that due to the presence of second-order derivatives of velocity fields in Eq. \ref{Eq2}, choosing an activation function with a well-defined second-order derivative is essential. The hyperbolic tangent activation function satisfies this criterion. Audiences interested in details of PIPN may refer to \citet{KASHEFI2022111510}. 

The residuals of conservation of mass ($r^{\text{continuity}}$), conservation of momentum in the $x-$direction ($r^{\text{momentum}_x}$) and in the $y-$direction ($r^{\text{momentum}_y}$), no slip boundary condition of the velocity ($r^{\text{velocity}_{\text{wall}}}$), sparse observations of the velocity field ($r^{\text{velocity}_{\text{obs}}}$) and pressure field ($r^{\text{pressure}_{\text{obs}}}$) are respectively written as

\begin{equation}
\label{Eq4}
r^{\text{continuity}} = \frac{1}{M_1} \sum_{k=1}^{M_1} \left (\frac{\delta \Tilde{u}'_k}{\delta \Tilde{x}_k} + \frac{\delta \Tilde{v}'_k}{\delta \Tilde{y}_k}  \right)^2,
\end{equation}

\begin{eqnarray}
\label{Eq5}
    r^{\text{momentum}_x} =
    \frac{1}{M_1} \sum_{k=1}^{M_1}
    \left ( \frac{\delta \Tilde{p}'_k}{\delta \Tilde{x}_k}
    -\Tilde{\mu} \left(\frac{\delta}{\delta \Tilde{x}_k} \left(\frac{\delta \Tilde{u}'_k}{\delta \Tilde{x}_k} \right) + \frac{\delta}{\delta \Tilde{y}_k}\left(\frac{\delta \Tilde{u}'_k}{\delta \Tilde{y}_k} \right) \right)  \right)^2,
\end{eqnarray}

\begin{eqnarray}
\label{Eq6}
r^{\text{momentum}_y} =
\frac{1}{M_1} \sum_{k=1}^{M_1}
\left ( \frac{\delta \Tilde{p}'_k}{\delta \Tilde{y}_k}-\Tilde{\mu} \left(\frac{\delta}{\delta \Tilde{x}_k} \left(\frac{\delta \Tilde{v}'_k}{\delta \Tilde{x}_k} \right) + \frac{\delta}{\delta \Tilde{y}_k}\left(\frac{\delta \Tilde{v}'_k}{\delta \Tilde{y}_k} \right) \right) \right)^2,
\end{eqnarray}

\begin{eqnarray}
\label{Eq7}
r^{\text{velocity}_{\text{wall}}} = \frac{1}{M_2} \sum_{k=1}^{M_2} \left (\left(\Tilde{u}'_k - 0 \right)^2 + \left(\Tilde{v}'_k - 0 \right)^2  \right),
\end{eqnarray}
\begin{eqnarray}
\label{Eq8}
r^{\text{velocity}_{\text{obs}}} = \frac{1}{M_3} \sum_{k=1}^{M_3} \left (\left(\Tilde{u}'_k - \Tilde{u}_k \right)^2 + \left(\Tilde{v}'_k - \Tilde{v}_k \right)^2  \right),
\end{eqnarray}
\begin{eqnarray}
\label{Eq9}
r^{\text{pressure}_{\text{obs}}} = \frac{1}{M_3} \sum_{k=1}^{M_3} \left (\Tilde{p}'_k - p_k \right)^2,
\end{eqnarray}
where $\delta$ stands for the automatic differentiation operator in the TensorFlow software \citep{abadi2016tensorflow}. The number of interior points, points located on wall boundaries, and virtual sensors measuring velocity and pressure values are respectively indicated by $M_1$, $M_2$, and $M_3$. Note that $M_1 + M_2=N$.

We normalize the output of the velocity and pressure fields because the output of the hyperbolic activation function (see Eq. \ref{S4}) only covers the range of [$-$1, 1]. The scaled ground truth velocity and pressure values are shown by ($\Tilde{u}$,$\Tilde{v}$,$\Tilde{p}$), while the predicted velocity and pressure fields by PIPN are denoted by ($\Tilde{u}'$,$\Tilde{v}'$,$\Tilde{p}'$). Additionally, the spatial coordinates of $x$ and $y$ (as the PIPN input) are scaled in the range of [$-$1, 1] and are shown by $\Tilde{x}$ and $\Tilde{y}$, respectively. Moreover, $\Tilde{\mu}$ is the scaled viscosity. In this sense, the final form of the loss function ($\mathcal{J}$) is determined as 

\begin{equation}
\label{Eq10}
\begin{split}
\mathcal{J}= \lambda_1 r^{\text{continuity}} + \lambda_2 r^{\text{momentum}_x} + \lambda_3 r^{\text{momentum}_y}
+ \lambda_4 r^{\text{velocity}_{\text{wall}}} 
\\
+\lambda_5 r^{\text{velocity}_{\text{obs}}} + \lambda_6 r^{\text{pressure}_{\text{obs}}},
\end{split}
\end{equation}
where $\lambda_i$ ($1\leq i \leq 6$) are the corresponding weights of each component of the loss function, while they take the inverse of the units of their associated residuals as their own unit. In this way, the loss function ($\mathcal{J}$) is unitless. These weights ($\lambda_i$; $1 \leq i \leq 6$) are, indeed, hyperparameters that need to be tuned for reaching the highest possible performance of PIPN. In this study, $\lambda_i$ are kept constant during training; however, one may use adaptive techniques for online tuning of $\lambda_i$ as the network is trained \citep{xiang2021self}.

\begin{figure*}
\label{Fig1}
\centering
\includegraphics[width=1.0\textwidth]{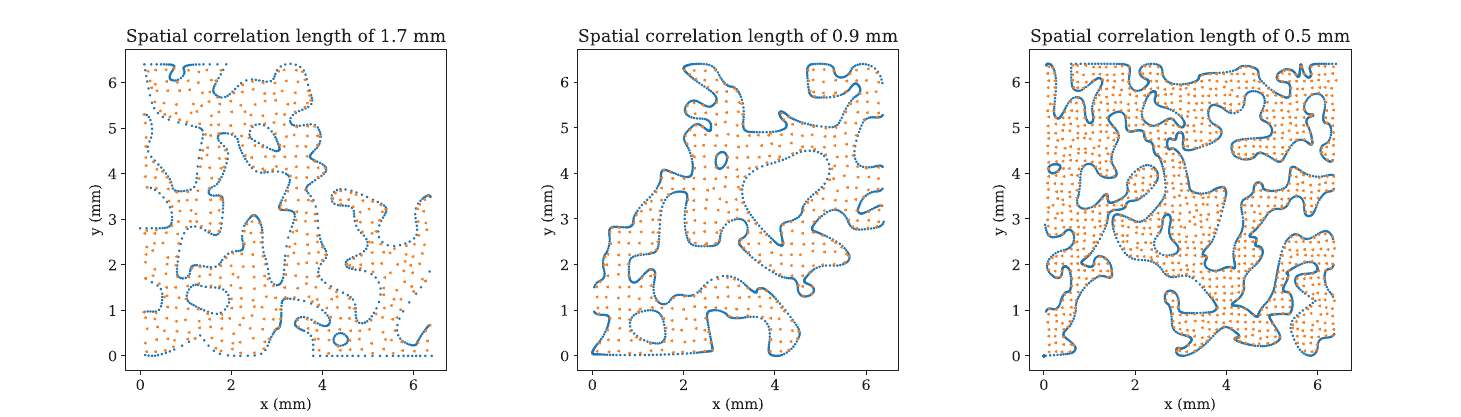}
\caption{Sensor locations for porous media with the spatial correlation lengths of 1.7 mm, 0.9 mm, and 0.5 mm}
\end{figure*}

\begin{figure*}
\label{FigPointCloud}
\centering
\includegraphics[width=1.0\textwidth]{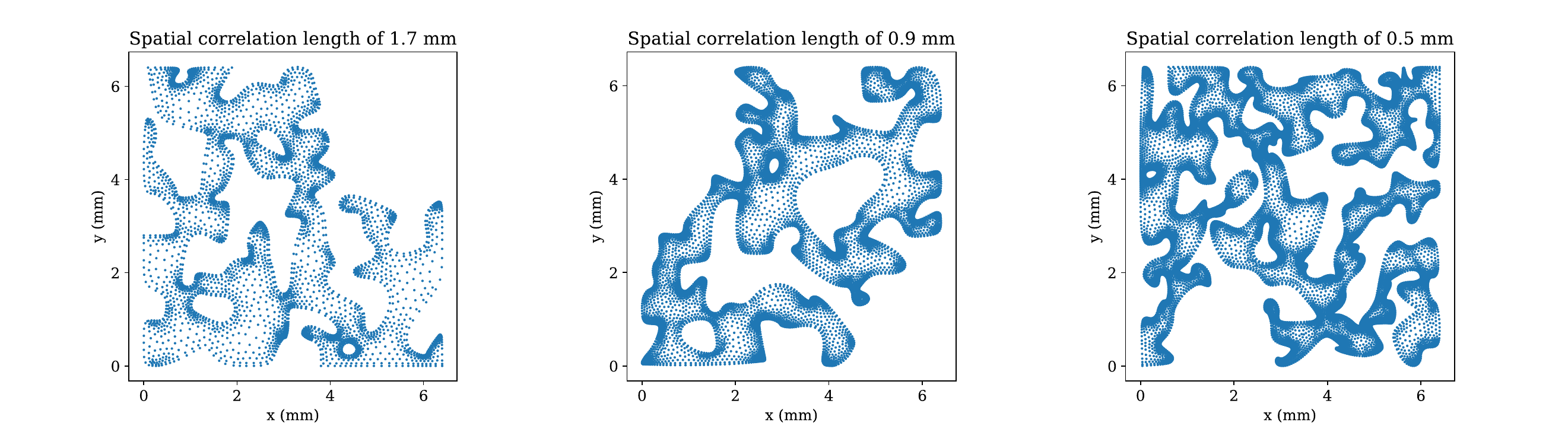}
\caption{Point clouds representing porous media with the spatial correlation lengths of 1.7 mm, 0.9 mm, and 0.5 mm}
\end{figure*}

\begin{table*}
\caption{Computational setup for the porous media considered in this study}
\centering
\renewcommand{\arraystretch}{2}
\begin{tabular}{l l l l}
\hline
Spatial correlation length ($l_c$) & 1.7 mm & 0.9 mm & 0.5 mm \\
\hline
Number of inquiry points ($N$) & 4231 & 8727 & 17661 \\
\hline
Number of virtual sensors ($M_3$) & 514 & 458 & 1120 \\
\hline
Rough percentage of observation (i.e., $M_3/N$) & 12\% & 5\% & 6\% \\
\hline
Average spatial distance of sensors ($d_s$) & 0.237 mm & 0.237 mm & 0.16 mm \\
\hline
\end{tabular}
\label{Tab1prime}
\end{table*}

\subsection{Computational setting}

For the numerical examples in the following, dynamic viscosity of $\mu = $ 0.001 Pa$\cdot$s, a pressure difference of $\Delta p= $ 0.1 Pa over a length of $L= $ 0.0064 m are set. We generate two-dimensional synthetic binary (pore-grain) media with the length ($L$) of 0.0064 m in both dimensions and three different spatial correlation lengths ($l_c$) of 0.0005 m, 0.0009 m, and 0.0017 m using the algorithm of truncated Gaussian simulation \citep{lantuejoul2001geostatistical,xu1993gtsim}. \textcolor{blue}{Practically, we first generate a two-dimensional array (64 by 64) of random numbers from the normal distribution with the mean parameter of 0.0 and the standard deviation of 1.0. Next, we filter the array using a two-dimensional Gaussian smoothing kernel with the standard deviation of 2.0 with a filter size equivalent to a desired spatial correlation length (e.g., 5, 9, or 17). Finally, we make the resulting array binary using a threshold such that the porosity falls in the range of [0.25, 0.40]. For these purposes, we use the MATLAB software. Afterward, we convert the resulting image (i.e., the two-dimensional array) to the standard tessellation language (STL) format such that it becomes readable by the COMSOL software.} By decreasing the spatial correlation length, the geometry of a porous medium becomes more complicated and the number of points representing the space of the corresponding porous medium in the point cloud likewise increases, imposing higher computational costs on the PIPN system. In this sense, the PIPN capability is validated for a variety of complexity levels.

Table \ref{Tab1prime} provides a fraction of sensor points to the total number of point cloud points (${M_3}/N$) as well as the average spatial distance of sensors from each other ($d_s$). Note that the sensor locations (see Fig. 2) are sparser than the point cloud (see Fig. 3) inputs of PIPN. Additionally, the point cloud (see Fig. 3) has a spatially varying density of points with denser points at boundaries and narrow pore throats.

Generally speaking, since the flow at the wall moves with zero velocity (i.e., no slip condition), $r^{\text{velocity}_{\text{wall}}}$ must have less weights compared to other residuals; otherwise, PIPN converges to the superficial solution of zero everywhere for all the fields. In this sense, for the porous media under investigation in this study, we set $\lambda_1= 100$ s, $\lambda_2=100$ m$^3$/N,
$\lambda_3=100$ m$^3$/N,
$\lambda_5=100$ s/m,
$\lambda_6=100$ m$^2$/N, and $\lambda_4=1$ s/m.

We use the Adam optimizer \citep{kingma2014adam} for training the PIPN configuration and set its associated hyperparameters as follows: $\beta_1=$ 0.9, $\beta_2=$ 0.999, and $\hat{\epsilon} = 10^{-6}$. One may  refer to \citet{kingma2014adam} for the definition of these hyperparameters. Furthermore, a constant learning rate of $\alpha=$ 0.0003 is chosen for all the porous media considered in this research letter. Training of PIPN is executed until satisfying the condition of $\mathcal{J} \leq 0.0025$. Machine learning computations are executed on a TESLA V100 graphic card with a memory clock rate of 1.38 GHz. \textcolor{blue}{To validate predictions of PIPN and generate sparse labeled data at virtual sensor locations, we employ the COMSOL software (see e.g., \citep{pirnia2019icy,jafari2023extended,azad2016comsol,shi2022comsol}) to solve Eqs. \ref{Eq1}--\ref{Eq2} using a finite element method (see e.g., \citep{kashefi2018finite,kashefi2020coarse}).} Alternatively, sparse observations can be practically obtained by lab experimental techniques \citep{bultreys2022x,sabbagh2020micro,karlsons2022integrating}. Additionally, grid vertices of the generated finite element meshes are taken to construct point clouds as the input of PIPN.

Finally, we address two points. First, PIPN was primarily designed for predicting the solutions of desired PDEs on multiple sets of irregular geometries, simultaneously. In this research; however, we employ PIPN for predicting the porous medium flows on a single geometry. Conceptually, one may even use regular PINNs for this purpose. Nevertheless, \citet{KASHEFI2022111510} showed that PIPN was more stable and required fewer inquiry points ($N$) even for training on a single geometry, compared to a regular PINN. These two features reduce computational costs by default. One may refer to Sect. 4.1.6 of \citet{KASHEFI2022111510} for a comprehensive comparison between PIPN and regular PINNs. Second, the size of PIPN is scalable and can be conveniently adjusted depending on the size of inquiry points ($N$). In this study, for example, we use a relatively larger PIPN for porous media with smaller spatial correlation lengths.

\begin{figure*}
\label{Fig2}
\centering
\includegraphics[width=1.0\textwidth]{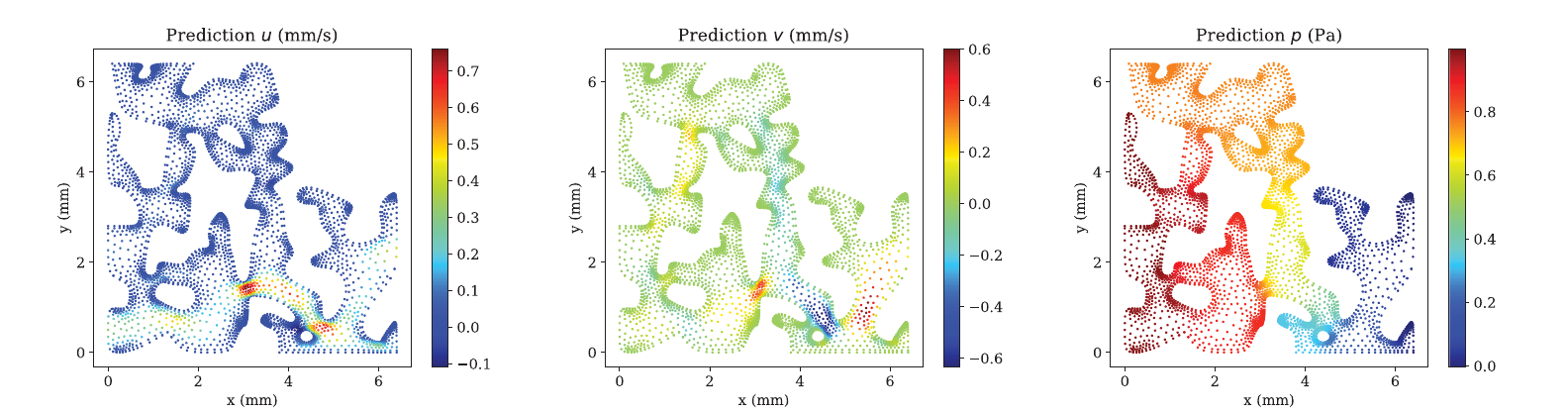}
\includegraphics[width=1.0\textwidth]{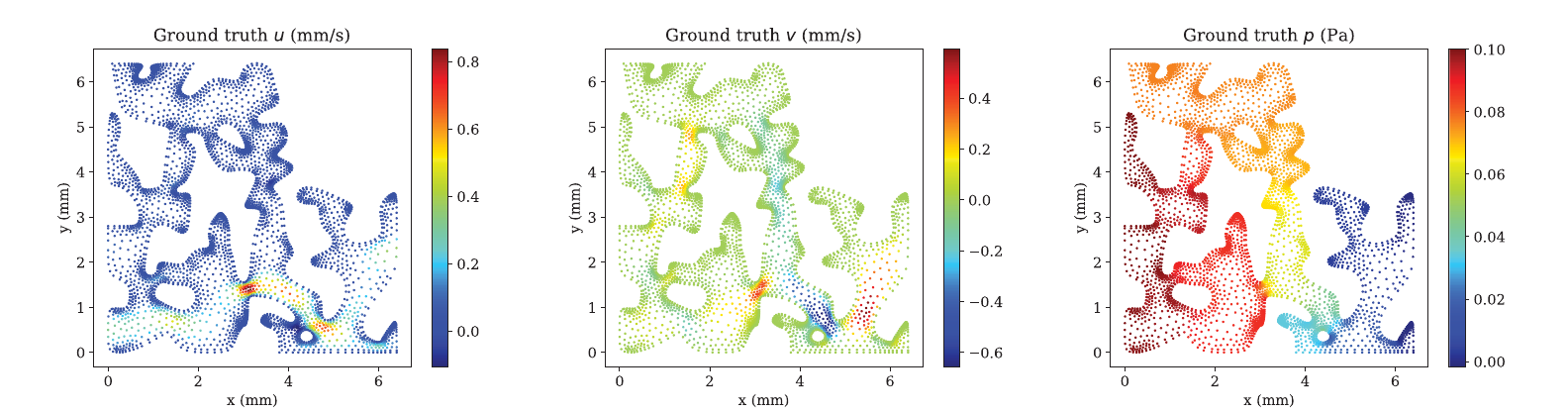}
\includegraphics[width=1.0\textwidth]{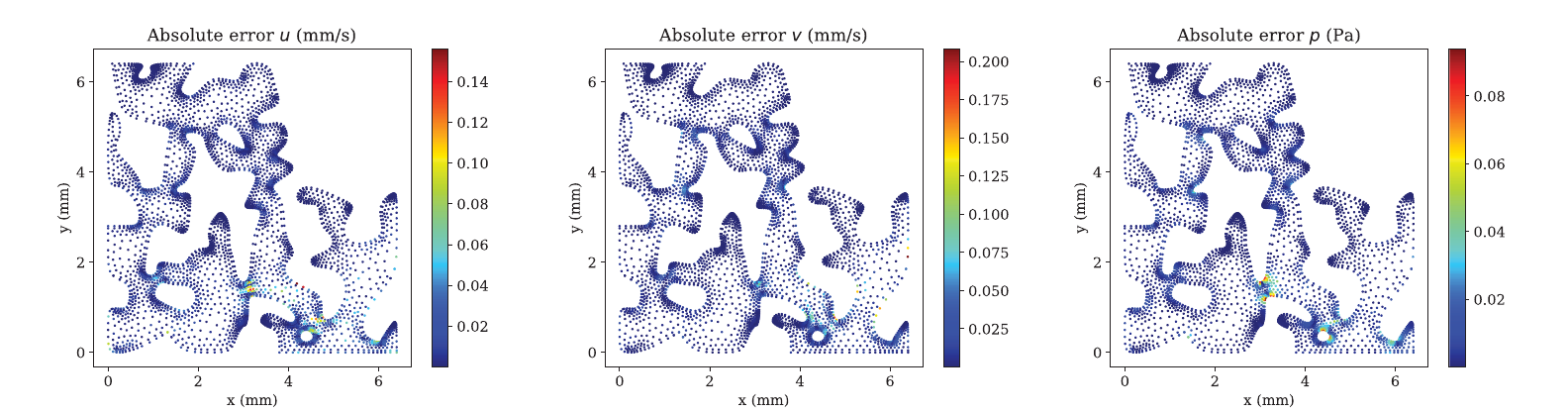}
\caption{\textcolor{blue}{ Comparison between the ground truth by COMSOL and the prediction by PIPN for the velocity and pressure fields of the porous medium with the spatial correlation length of $l_c=$ 1.7 mm}}
\end{figure*}

\begin{figure*}
\label{Fig3}
\centering
\includegraphics[width=1.0\textwidth]{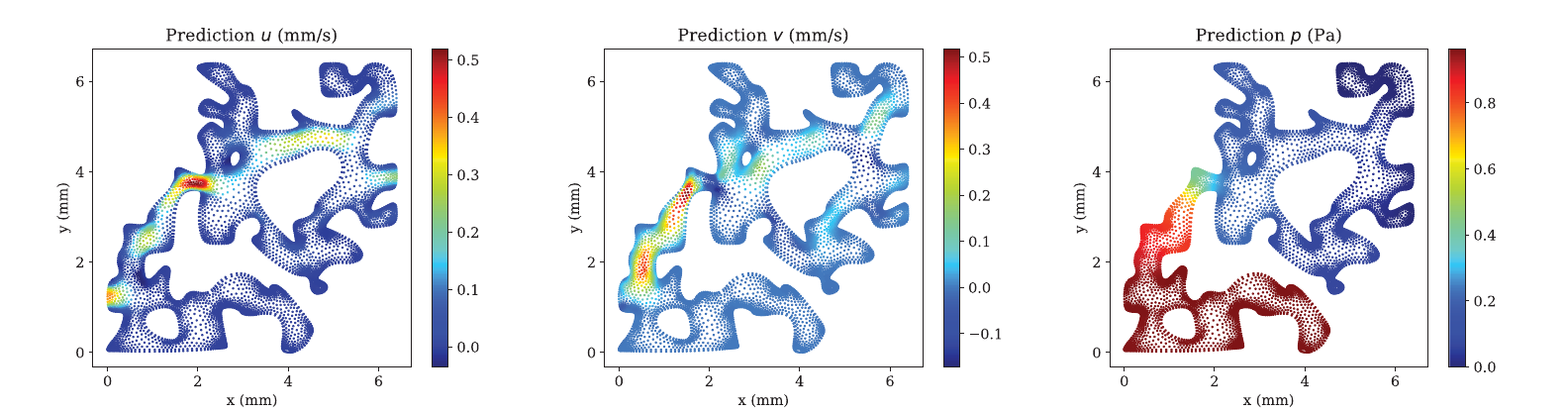}
\includegraphics[width=1.0\textwidth]{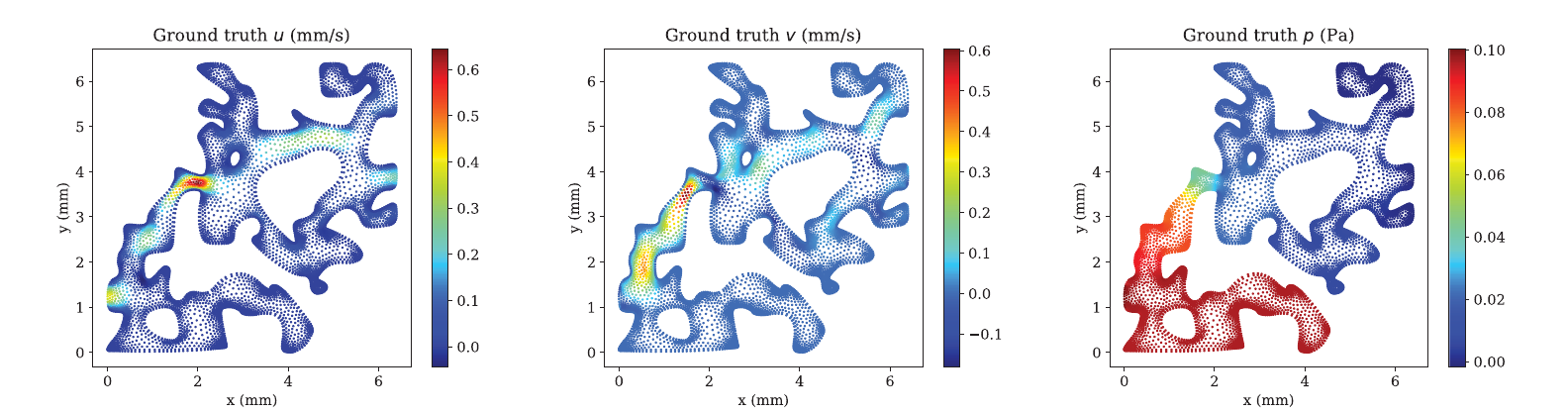}
\includegraphics[width=1.0\textwidth]{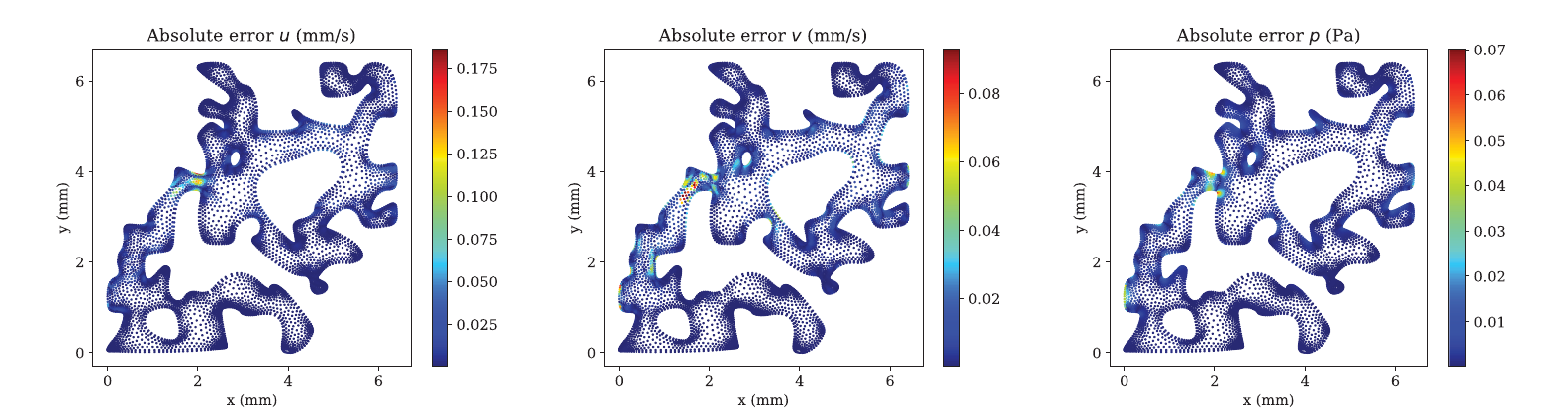}
\caption{\textcolor{blue}{ Comparison between the ground truth by COMSOL and the prediction by PIPN for the velocity and pressure fields of the porous medium with the spatial correlation length of $l_c=$ 0.9 mm}}
\end{figure*}

\section{Results and discussion}
\subsection{General analysis}

A visual comparison between the ground truth and the PIPN prediction for the velocity and pressure fields is made for the porous media with spatial correlation lengths ($l_c$) of 1.7 mm, 0.9 mm, and 0.5 mm respectively in Figures 4, 5, and 6. All in all, a good agreement between the prediction and ground truth at the point cloud locations is observed. In all three cases, the maximum local errors for the velocity fields happen in the narrowest bottlenecks, where the flow accelerates. \textcolor{red}{This observation highlights the importance of conducting precise measurements with a sufficient number of sensors in these locations. Technically; however, this is possible only if the space in the narrowest channels is yet large enough. Presumably, if this was not the case for a specific situation, we would suggest a potential technique to handle this challenge as follows. After obtaining the velocity field predicted by PIPN, one may use this predicted solution as an initial guess for a numerical classical solver (e.g., COMSOL) and execute the solver. Note that because the prediction by PIPN experiences a low level of errors (as shown in Figures 4--6) it might only take a few iterations for the solver to converge with an improvement in the accuracy of the velocity fields in the narrowest bottlenecks.}

The order of accuracy of the velocity fields for all three porous media is approximately the same due to the fact that PIPN is enforced to satisfy the same criterion (i.e., $\mathcal{J} \leq 0.0025$). Nevertheless, the porous media with shorter spatial correlation lengths ($l_c$) requires a higher number of iterations due to having a higher number of inquiries and more complicated geometry, as listed in Table \ref{Tab1}. Note that the wall time consumed per epoch is approximately equal for all three porous media under investigation. This is because the TensorFlow \cite{abadi2016tensorflow} software constructs the computation graph only once at the beginning of the training and uses it in the rest of the training for calculations. In fact, although the computation graph becomes more complex by decreasing the spatial correlation length, the computation per iteration (i.e., per epoch) only slightly (a few seconds) enhances, as TensorFlow \cite{abadi2016tensorflow} simply updates numbers in the previously constructed graph to calculate the loss value.

To validate more precisely the PIPN performance, we tabulate the relative pointwise error ($L^2$ norm) of the PIPN prediction in Table \ref{Tab1}. Accordingly, the relative error of the velocity fields is approximately in the range of 15\% to 19\%. In all the cases, the predicted velocity fields experience a higher level of errors compared to the pressure one due to the fact that the pressure simply linearly decreases in the $x-$direction, whereas the velocity pattern is more complicated. Additionally, the velocity vector is involved in both the mass and momentum balances, while the pressure only plays a role in the momentum equation (Eq. \ref{Eq2}). According to Table \ref{Tab1}, the relative error of the obtained permeability (see Eq. \ref{Eq3}) of the porous media as a result of the predicted velocity field is less than 5\% for all three cases. Note that the average spatial distance of sensors from each other ($d_s$) indicates the level of information that we observe for each porous medium. For the porous media with $l_c= $ 1.7 mm and $l_c= $ 0.9 mm, $d_s= $ 0.237 mm, while for the porous medium with $l_c= $ 0.5 mm, $d_s= $ 0.16 mm. Hence, we observe more information for the medium with $l_c= $ 0.5 mm compared to the two others. As mentioned earlier, these reported $d_s$ are the maximum possible distance for satisfying the convergence criterion in PIPN (i.e., $\mathcal{J} \leq 0.0025$).


\begin{figure*}
\label{Fig4}
\centering
\includegraphics[width=1.0\textwidth]{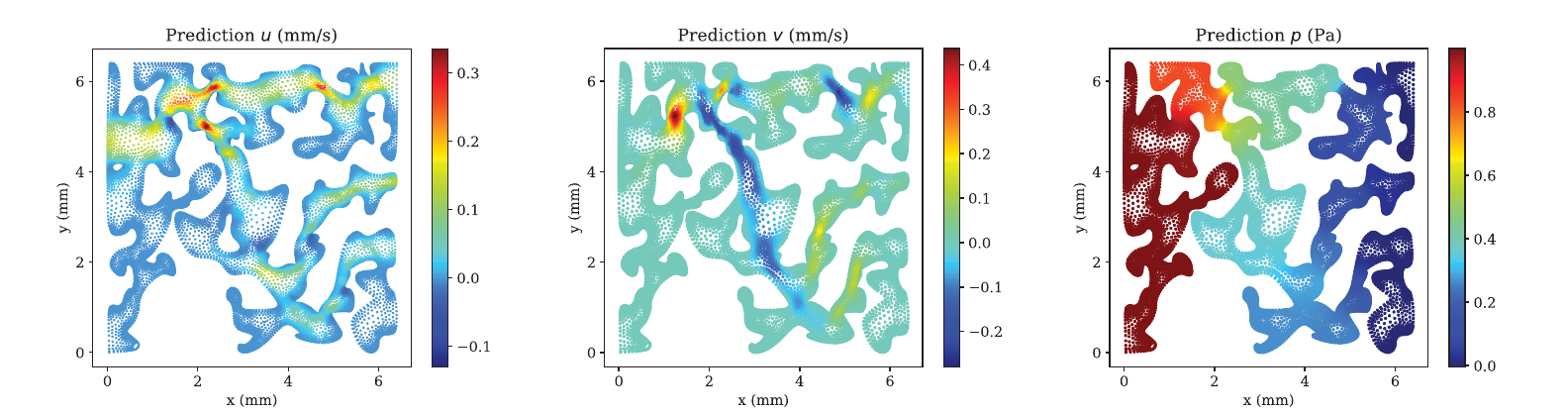}
\includegraphics[width=1.0\textwidth]{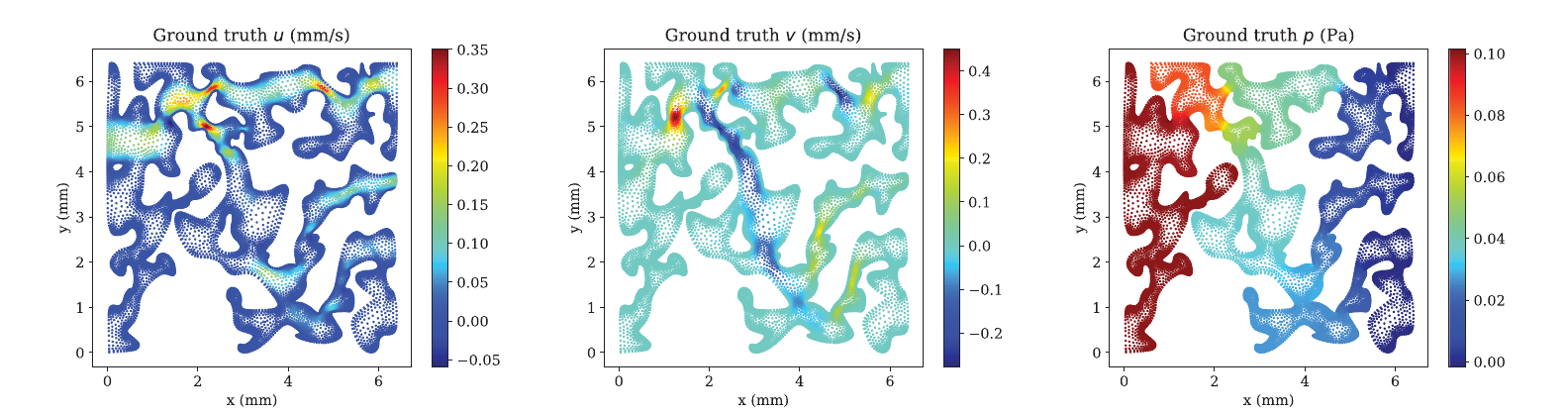}
\includegraphics[width=1.0\textwidth]{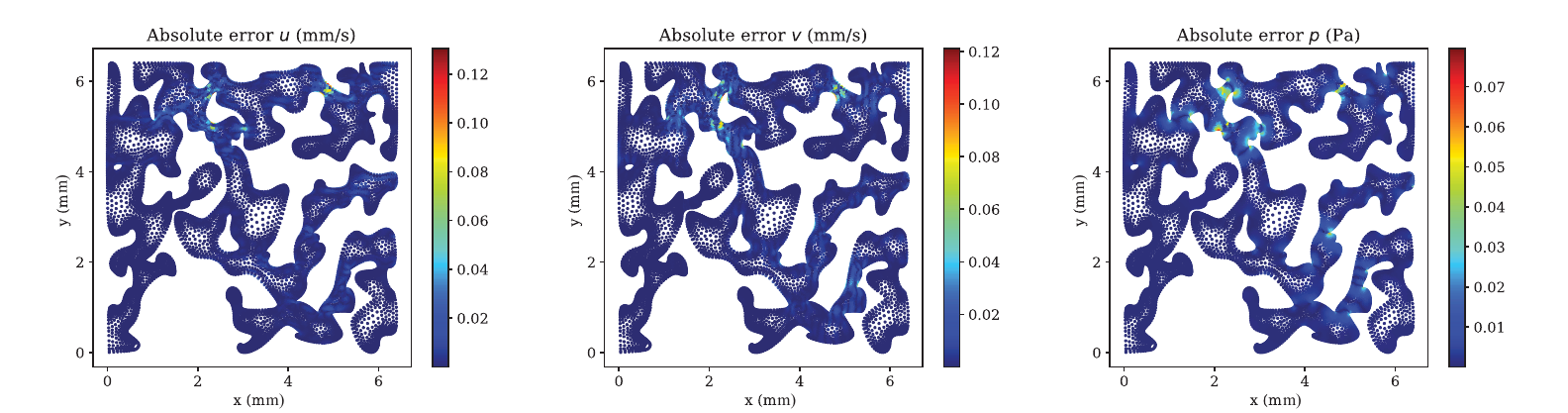}
\caption{\textcolor{blue}{ Comparison between the ground truth by COMSOL and the prediction by PIPN for the velocity and pressure fields of the porous medium with the spatial correlation length of $l_c=$ 0.5 mm}}
\end{figure*}

\subsection{Effect of sparse pressure observations}

Next, we investigate the influence of the pressure observation on the prediction accuracy of PIPN. This investigation is important and critical because sparse observations of the pressure field may not be available for various reasons such as technical difficulties or expensive lab experiments. To this end, we modify the loss function by dropping the pressure observation residual $r^{\text{pressure}_{\text{obs}}}$, in the loss function yielding

\begin{equation}
\label{Eq11}
\begin{split}
    \mathcal{J}= \lambda_1 r^{\text{continuity}} + \lambda_2 r^{\text{momentum}_x} + \lambda_3 r^{\text{momentum}_y}
    + \lambda_4 r^{\text{velocity}_{\text{wall}}}
    \\
    + \lambda_5 r^{\text{velocity}_{\text{obs}}}.
\end{split}
\end{equation}
Similar to the previous subsection, we set $\lambda_1= 100$ s, $\lambda_2=100$ m$^3$/N,
$\lambda_3=100$ m$^3$/N,
$\lambda_5=100$ s/m, and $\lambda_4=1$ s/m. We carry out this machine learning experiment for the porous media with the spatial correlation length of $l_c = 1.7$ mm, $l_c = 0.9$ mm, and $l_c = 0.5$ mm; and the outcomes are tabulated in Table \ref{TabPressure}.

For example, let us discuss the results obtained for the porous medium with the spatial correlation length of $l_c = 1.7$ mm. As a consequence of this modification, the relative error of the predicted velocity field in the $x$ and $y$ directions and the predicted pressure field become 1.43444E$-$1, 1.71549E$-$1, and 1.25499, respectively. We observe that although the relative error of the predicted pressure field increases enormously (by 667.024\%), the accuracy of the predicted velocity field remains approximately unchanged. A similar observation has been reported by \citet{KASHEFI2022111510} for the natural convection problem. Reasons for this evidence have been articulated in detail both from applied mathematics and machine learning perspectives in Sect. 4.2.1 of \citet{KASHEFI2022111510}. But in a nutshell, it can be explained as follows. PIPN is able to preserve the accuracy of the pressure gradient in the absence of pressure observations in the Stokes equations (Eqs. \ref{Eq1}--\ref{Eq2}). This feature of PIPN relies on the fact that pressure is an implicit variable in the steady Stokes equations \citep{timmermans1996approximate}. Similar information can be observed and realized for the porous media with the spatial correlation length of $l_c = 0.9$ mm and $l_c = 0.5$ mm, as can be inferred in Table \ref{TabPressure}.

By comparing the information presented in Table \ref{Tab1} and Table \ref{TabPressure}, the number of required iterations (i.e., epochs) for convergence satisfaction decreases for the current test cases. This is because the PIPN loss function becomes less restricted by dropping the pressure observation residual ($r^{\text{pressure}_{\text{obs}}}$) from the loss function. Additionally, since the pressure term is omitted from the PIPN loss function, the computation graph associated with the loss function becomes slightly less complex, and consequently, the wall time expended per iteration (i.e., epoch) slightly lessens.

All in all, the outputs discussed in this subsection show that the PIPN methodology successfully reliably predicts the velocity fields of the porous media even in the absence of the sparse pressure data.


\begin{table*}
\caption{Error analysis of the velocity, pressure, and permeability predicted by PIPN for porous
media with three different spatial correlation lengths; The $L^2$ norm is indicated by $||\cdots||$ and the absolute norm is shown by $| \cdots|$.}
\centering
\renewcommand{\arraystretch}{2}
\begin{tabular}{l l l l}
\hline
Spatial correlation length ($l_c$) & 1.7 mm & 0.9 mm & 0.5 mm \\
\hline
$\frac{||u-\Tilde{u}||}{||u||}$  & 1.51927E$-$1 & 1.54239E$-$1 & 1.79362E$-$1 \\
\hline
$\frac{||v-\Tilde{v}||}{||v||}$  &  1.84293E$-$1 & 1.47009E$-$1 & 1.79883E$-$1 \\
\hline
$\frac{||p-\Tilde{p}||}{||p||}$  & 1.63618E$-$2 & 7.34894E$-$3 & 9.96776E$-$3 \\
\hline
$\frac{|\mathcal{K}-\Tilde{\mathcal{K}}|}{|\mathcal{K}|}$  & 4.21557E$-$2 & 8.88220E$-$3 & 3.29516E$-$2 \\
\hline
Number of iterations (i.e., epochs) & 113141 & 149819 & 162145 \\
\hline
Wall time per iteration (i.e., epoch) & 31.4 s & 32.3 s &  33.6 s \\
\hline
\end{tabular}
\label{Tab1}
\end{table*}


\begin{table*}
\caption{Investigation of the effect of the absence of the pressure measurements in the loss function (see Eq. \ref{Eq11}) and the resulting relative errors of the velocity, pressure, and permeability predicted by PIPN for porous
media with three different spatial correlation lengths; The $L^2$ norm is indicated by $||\cdots||$.}
\centering
\renewcommand{\arraystretch}{2}
\begin{tabular}{l l l l}
\hline
Spatial correlation length ($l_c$) & 1.7 mm & 0.9 mm & 0.5 mm \\
\hline
$\frac{||u-\Tilde{u}||}{||u||}$  & 1.43444E$-$1 & 1.99840E$-$1 & 1.99135E$-$1 \\
\hline
$\frac{||v-\Tilde{v}||}{||v||}$  &  1.71549E$-$1 & 1.87410E$-$1 & 1.84841E$-$1 \\
\hline
$\frac{||p-\Tilde{p}||}{||p||}$  & 1.25499 & 7.05858E$-$1 & 6.64332E$-$1 \\
\hline
Number of iterations (i.e., epochs) & 15519 & 32601 & 41426 \\
\hline
Wall time per iteration (i.e., epoch) & 30.9 s & 31.5 s & 33.4 s \\
\hline
\end{tabular}
\label{TabPressure}
\end{table*}


\subsection{Effect of noisy data}

Sensor measurements are usually polluted by noises. To mimic this scenario, for instance, we add 5\% random Gaussian noise to the velocity and pressure observations at all sensor locations for the porous media with the spatial correlation length of $l_c=1.7$ mm, $l_c=$ 0.9 mm, and $l_c=$ 0.5 mm. The results and error analysis are tabulated in Table \ref{TabNoise}.

As can be realized from the information of Table \ref{TabNoise}, the error of the predicted velocity and pressure field increases as a result of adding noise to the sparse data. For example, considering the porous medium with the spatial correlation length of $l_c = 1.7$ mm, the relative pointwise error ($L^2$ norm) of the predicted $u$ component of the velocity, $v$ component of the velocity, and pressure fields become respectively 2.04158E$-$1, 2.16941E$-$1, and 2.58180E$-$2, indicating 16.278\%, 6.983\%, and 25.110\% increase compared to the results obtained with noise-free data (see Table \ref{Tab1} and Table \ref{TabNoise}).

Due to the presence of noise, the PIPN loss function is unable to satisfy the convergence criterion (i.e., $\mathcal{J} \leq 0.0025$) and this is why the PIPN predictions come with a higher level of errors compared to the deep learning experiment with the noise-free data (see Table \ref{Tab1} and Table \ref{TabNoise}). In practice, we continue the training procedure until the loss value reaches a plateau. The corresponding number of iterations for each porous medium is listed in Table \ref{TabNoise}. By comparing Table \ref{TabNoise} and Table \ref{Tab1}, it is realized that the number of iterations for reaching even a plateau (in the case of the noisy data) is greater than the number of required iterations for satisfying the convergence criterion (in case of the clean data), demonstrating that the noisy data demands more computational costs for the PIPN deep learning solver. Having said that, the relative pointwise error ($L^2$ norm) is less than approximately 22\% for the velocity and pressure variables for all the porous media mentioned in Table \ref{TabNoise}. By and large, it is concluded that the PIPN methodology is robust even in the presence of noisy sensor data observed in the porous media.



\begin{table*}
\caption{Error analysis of the velocity, pressure, and permeability predicted by PIPN for porous
media with three different spatial correlation lengths when the observed data is polluted with 5\% Gaussian noise; The $L^2$ norm is indicated by $||\cdots||$.}
\centering
\renewcommand{\arraystretch}{2}
\begin{tabular}{l l l l}
\hline
Spatial correlation length ($l_c$) & 1.7 mm & 0.9 mm & 0.5 mm \\
\hline
$\frac{||u-\Tilde{u}||}{||u||}$  & 1.76658E$-$1 & 2.04158E$-$1 & 2.03771E$-$1 \\
\hline
$\frac{||v-\Tilde{v}||}{||v||}$  &  1.97164E$-$1 & 2.16941E$-$1 & 1.88708E$-$1 \\
\hline
$\frac{||p-\Tilde{p}||}{||p||}$  & 2.04703E$-$2 & 2.58180E$-$2 & 2.30487E$-$2 \\
\hline
Number of iterations to reach a plateau  & 212401 & 240231 & 230171 \\
\hline
\end{tabular}
\label{TabNoise}
\end{table*}


\begin{figure*}
\label{FigDs}
\centering
\includegraphics[width=1.0\textwidth]{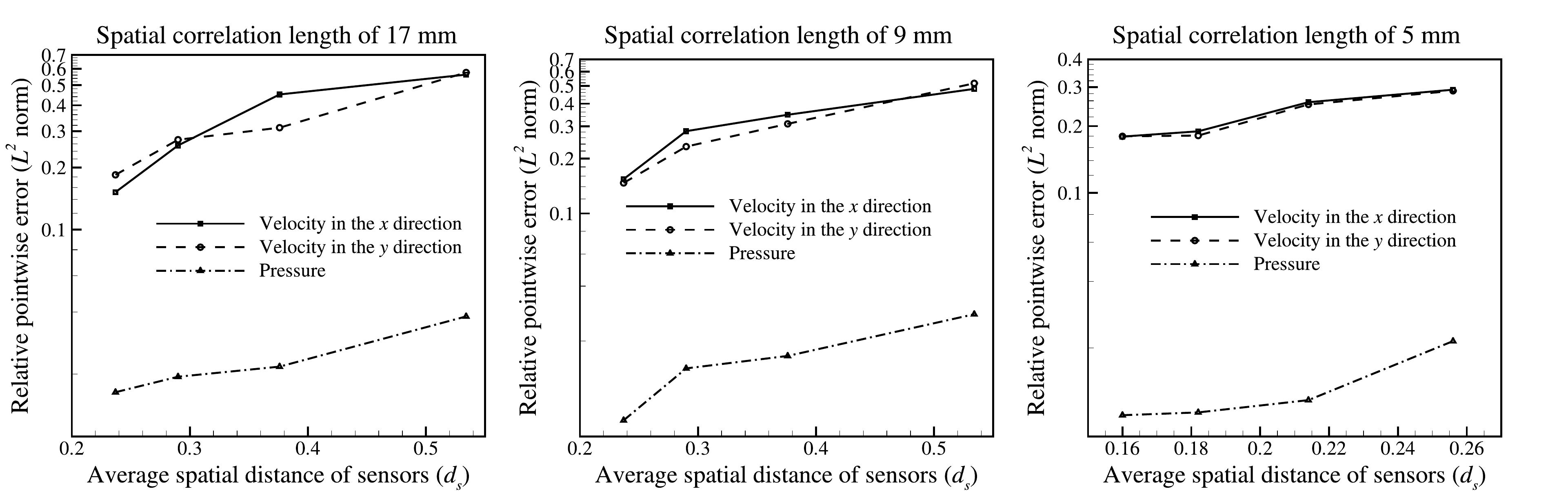}
\caption{Relative pointwise error ($L^2$ norm) as a function of the average spatial distance of sensors ($d_s$) for the porous media with the spatial correlation lengths ($l_c$) of 17 mm, 9 mm, and 5 mm}
\end{figure*}


\subsection{Effect of the average spatial distance of sensors}

In this section, the effect of the average spatial distance of sensors ($d_s$) on the PIPN prediction accuracy is investigated. To this end, we plot the relative pointwise error ($L^2$ norm) as a function of $d_s$ for the porous media with the spatial correlation length of $l_c=17$ mm, $l_c=9$ mm, and $l_c=5$ mm in Fig. 7. As can be observed in Fig. 7, by increasing $d_s$ (i.e., decreasing the number of sensors), the relative pointwise error ($L^2$ norm) increases. For instance, by increasing $d_s$ from 0.237 mm to 0.534 mm in the porous medium with the spatial correlation length ($l_c$) of 17 mm, the relative pointwise error ($L^2$ norm) of the velocity in the $x$ direction, the velocity in the $y$ direction, and the pressure field approximately increases by 270\%, 212\%, and 132\%. In fact, the results shown plotted in Fig. 7 reflects the importance of the role of sparse observations in the accuracy of the PIPN outputs.

Now, the question is by increasing $d_s$ how the prediction error spatially grows in the porous medium domain? To answer this question, we show the absolute pointwise errors of the velocity and pressure fields predicted by the PIPN methodology for different values of $d_s$, for example, for the porous media with the spatial correlation length ($l_c$) of 17 mm and 9 mm, respectively, in Fig. 8 and Fig. 9. Furthermore, the sensor locations associated with each $d_s$ are exhibited in Fig. 8 and Fig. 9. In all the cases, maximum local errors happen in areas where the fluid flows, specifically with higher accelerations. By increasing $d_s$ (i.e., decreasing sensor numbers), both the absolute value of the pointwise errors as well as the spatial areas polluted with significant errors in the porous media increase. After performing this machine learning experiment, at the first glance, it seems that one should set more sensors in the area where fluid flows and accelerates. However, the issue with this strategy is that these specific areas are unknown to us before executing the PIPN solver. Consequently, it is reasonable to spread the sensors at equal distances from each other throughout the entire space of a porous medium of interest.


\begin{figure*}
\label{ErrorDist17}
\centering
\includegraphics[width=1.0\textwidth]{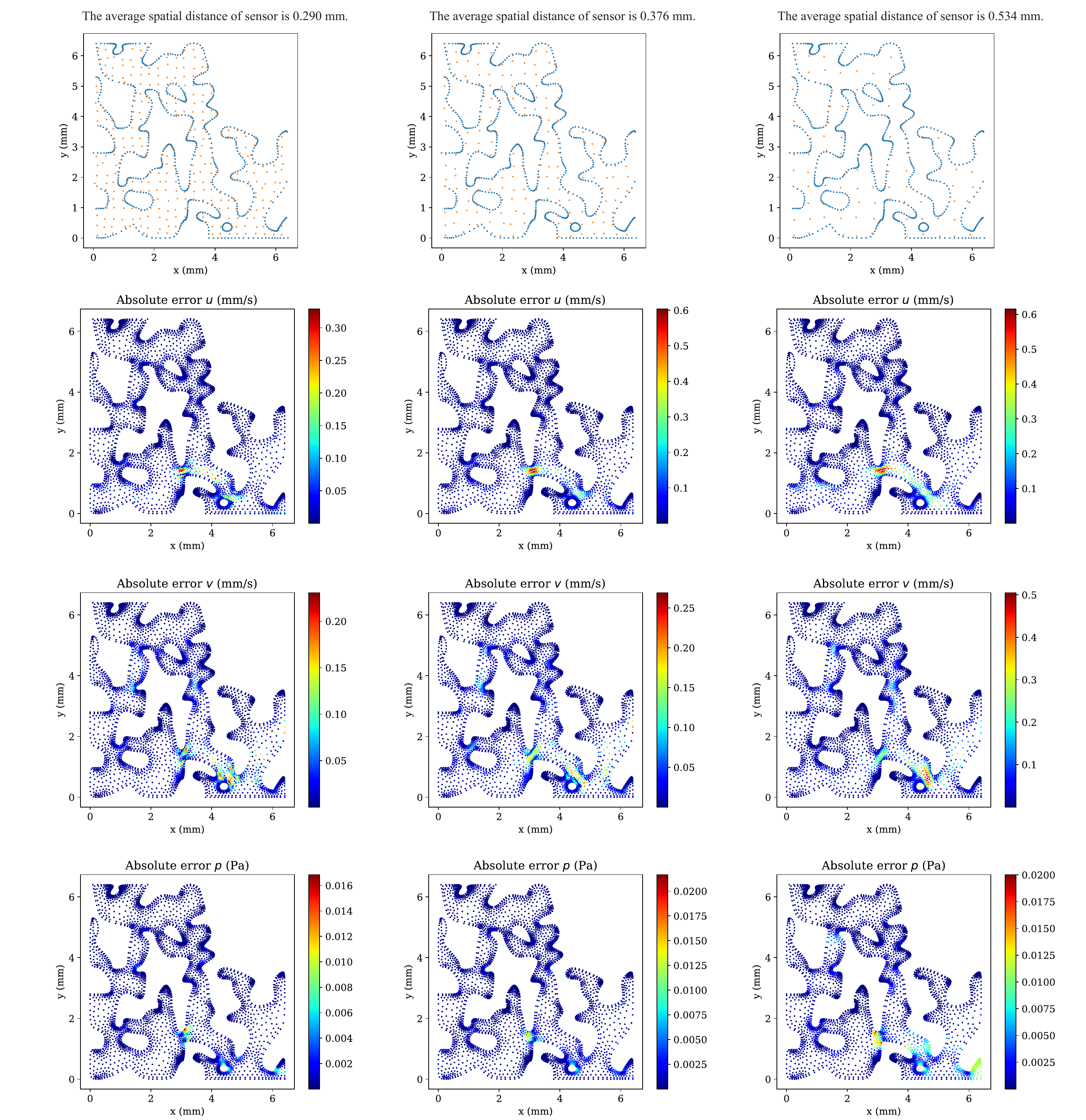}
\caption{\textcolor{blue}{Absolute pointwise error of the velocity and pressure fields predicted by PIPN for the porous medium with the spatial correlation length ($l_c$) of 17 mm when the average spatial distance of sensors ($d_s$) is 0.290 mm, 0.376 mm, and 0.534 mm. The solution by COMSOL is considered as the reference ground truth to compute the absolute pointwise error.}}
\end{figure*}


\begin{figure*}
\label{ErrorDist9}
\centering
\includegraphics[width=1.0\textwidth]{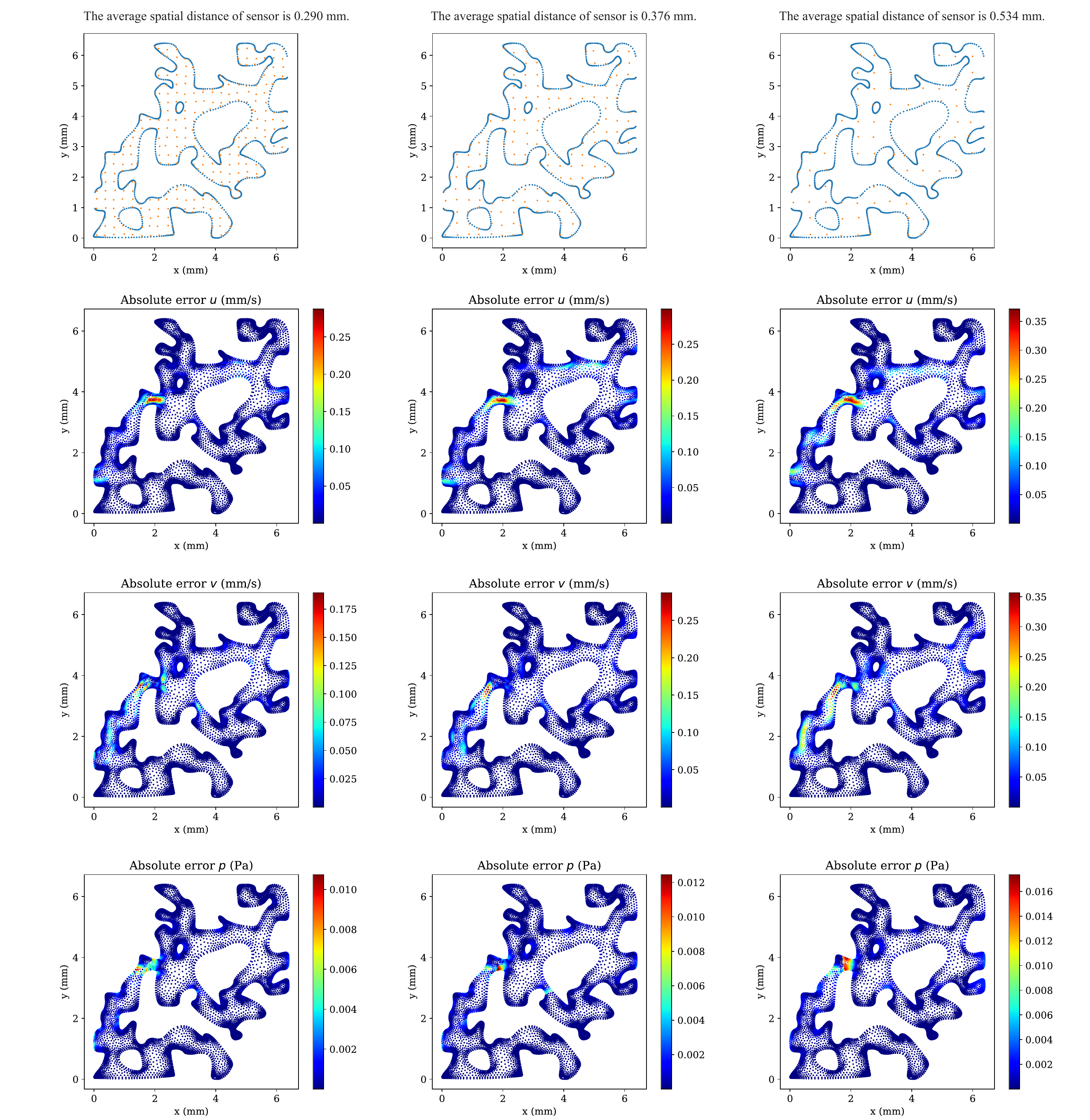}
\caption{\textcolor{blue}{Absolute pointwise error of the velocity and pressure fields predicted by PIPN for the porous medium with the spatial correlation length ($l_c$) of 9 mm when the average spatial distance of sensors ($d_s$) is 0.290 mm, 0.376 mm, and 0.534 mm. The solution by COMSOL is considered as the reference ground truth to compute the absolute pointwise error.}}
\end{figure*}


\section{Conclusions and future directions}
In this research letter, we applied PIPN \citep{KASHEFI2022111510} as an innovative physics-informed deep learning strategy for the prediction of the velocity and pressure fields of two-dimensional steady incompressible flows in porous media, while only a sparse scattered set of labeled data were observed. 
Using PIPN, first, we lessened the required computational memories by not taking the grain spaces of porous media into the machine learning framework. Second, the point cloud allowed us to represent the geometry of pore spaces of porous media more realistically. Third, we had the freedom to vary the spatial resolution of pores spaces to optimize the computational costs. Specifically, the effect of noisy sensor data, pressure observations, and spatial correlation lengths was investigated through visual results and quantitative error analysis.


\textcolor{red}{
One of our outlook projects is the extension of PIPN to three-dimensional and multiphase flows in porous media. More specifically, there are currently serious challenges for three-dimensional modeling (e.g., see \citep{saxena2017references}). By switching to a three-dimensional space, the number of points ($N$) in point clouds of porous media significantly increases and it leads to more complicated and lengthier computation graphs in PIPN, requiring larger GPU memories and longer wall clock time for running the PIPN platform. Additionally, it is conjectured that the accuracy of predictions by PIPN may be more sensitive to the number of sensors and the spatial distribution of sensors for the inverse problem in a three-dimensional space. One approach to overcome these barriers is to moderate the associated computational costs via parallel computing and domain decomposition techniques \citep{shukla2021parallel}.}


\section*{Author contributions}

\noindent
\textbf{Ali Kashefi:} Conceptualization, Methodology, Software, Formal analysis, Visualization, Writing - original draft, Writing - review \& editing

\noindent
\textbf{Tapan Mukerji:} Conceptualization, Methodology, Formal analysis, Writing - review \& editing, Project administration, Funding acquisition

\section*{Declaration of competing interest}
The authors declare that they have no known competing financial interests or personal relationships that could have appeared to influence the work reported in this paper.

\begin{acknowledgements}
Funding from Shell-Stanford collaborative project on Digital Rock Physics 2.0 is acknowledged for supporting this research project.
Moreover, we are thankful to the Stanford Research Computing Center for computational resources. \textcolor{blue}{Additionally, the authors would like to thank the reviewers for their beneficial comments and suggestions.}
\end{acknowledgements}

\section*{Computer Code Availability}
The software and data are available on the following GitHub repository:

\url{https://github.com/Ali-Stanford/PhysicsInformedPointNetPorousMedia}.
\begin{itemize}
\item Name of code: PIPN\_for\_Porous\_Media.py 
\item Developer and contact address: Ali Kashefi (kashefi@stanford.edu)
\item Year first available: 2022
\item Hardware required: Graphics Processing Unit (GPU) \textcolor{red}{with at least 48 GB of RAM}
\item Software required: TensorFlow; Python; Matplotlib; Numpy 
\item Program language: Python
\item Program size: 600 lines
\end{itemize}

\section*{Data availability}
The authors have shared the link to the data in the ``Computer Code Availability'' section.
\bibliographystyle{agu}   
\bibliography{template.bib}   

%
%

\end{document}